\documentclass[letterpaper, 10 pt, conference]{ieeeconf}  % Comment this line out if you need a4paper

\IEEEoverridecommandlockouts                              % This command is only needed if 
\usepackage{graphics} % for pdf, bitmapped graphics files
\usepackage{epsfig} % for postscript graphics files
\usepackage{mathptmx} % assumes new font selection scheme installed
\usepackage{times} % assumes new font selection scheme installed
\usepackage{amsmath} % assumes amsmath package installed
\usepackage{amssymb}  % assumes amsmath package installed
\usepackage{subcaption}
\usepackage{caption} % DO NOT CHANGE THIS AND DO NOT ADD ANY OPTIONS TO IT

\usepackage{ifthen}
\usepackage[ruled,vlined,linesnumbered]{algorithm2e}
\SetAlFnt{\scriptsize}

\title{\LARGE \bf
Distributed Multi-agent Navigation Based on Reciprocal Collision Avoidance and Locally Confined Multi-agent Path Finding
}

\author{Stepan Dergachev$^{1}$ and Konstantin Yakovlev$^{1}$% <-this % stops a space
\thanks{*This work was supported by RFBR, Grant 20-57-00011.}% <-this % stops a space
\thanks{$^{1}$Stepan Dergachev and Konstantin Yakovlev are with HSE University and Federal Research Center for Computer Science and Control of Russian Academy of Sciences. Moscow, Russia.}
\thanks{Corresponding author is Stepan Dergachev, \texttt{sadergachev@edu.hse.ru}.
}
}

\begin{document}

\newcommand\blfootnote[1]{%
  \begingroup
  \renewcommand\thefootnote{}\footnote{#1}%
  \addtocounter{footnote}{-1}%
  \endgroup
}

\maketitle
\thispagestyle{empty}
\pagestyle{empty}

%%%%%%%%%%%%%%%%%%%%%%%%%%%%%%%%%%%%%%%%%%%%%%%%%%%%%%%%%%%%%%%%%%%%%%%%%%%%%%%%
\begin{abstract}

Avoiding collisions is the core problem in multi-agent navigation. In decentralized settings, when agents have limited communication and sensory capabilities, collisions are typically avoided in a reactive fashion, relying on local observations/communications. Prominent collision avoidance techniques, e.g. \textsc{ORCA}, are computationally efficient and scale well to a large number of agents. However, in numerous scenarios, involving navigation through the tight passages or confined spaces, deadlocks are likely to occur due to the egoistic behaviour of the agents and as a result, the latter can not achieve their goals. To this end, we suggest an application of the locally confined  multi-agent path finding (MAPF) solvers that coordinate sub-groups of the agents that appear to be in a deadlock (to detect the latter we suggest a simple, yet efficient ad-hoc routine). We present a way to build a grid-based MAPF instance, typically required by modern MAPF solvers. We evaluate two of them in our experiments, i.e. \textsc{Push and Rotate} and a bounded-suboptimal version of \textsc{Conflict Based Search} (\textsc{ECBS}), and show that their inclusion into the navigation pipeline significantly increases the success rate, from 15\% to 99\% in certain cases.

\blfootnote{This is a preprint of the paper accepted to CASE'21}.

\end{abstract}

%%%%%%%%%%%%%%%%%%%%%%%%%%%%%%%%%%%%%%%%%%%%%%%%%%%%%%%%%%%%%%%%%%%%%%%%%%%%%%%%
\section{Introduction}

Multi-agent navigation is a challenging task arising in mobile robotics, video games, crowd simulation, etc. The approaches to solve this problem can be roughly divided into centralized and decentralized ones. Centralized approaches assume that there exists a (central) controller that possesses all the information about the agents and the environment and can communicate with the agents. The controller creates a joint collision-free plan and then lets the agents execute it. One of the main advantages of such an approach is strong theoretical guarantees.

\begin{figure}[t]
    \centering
    \includegraphics[width=0.75\columnwidth]{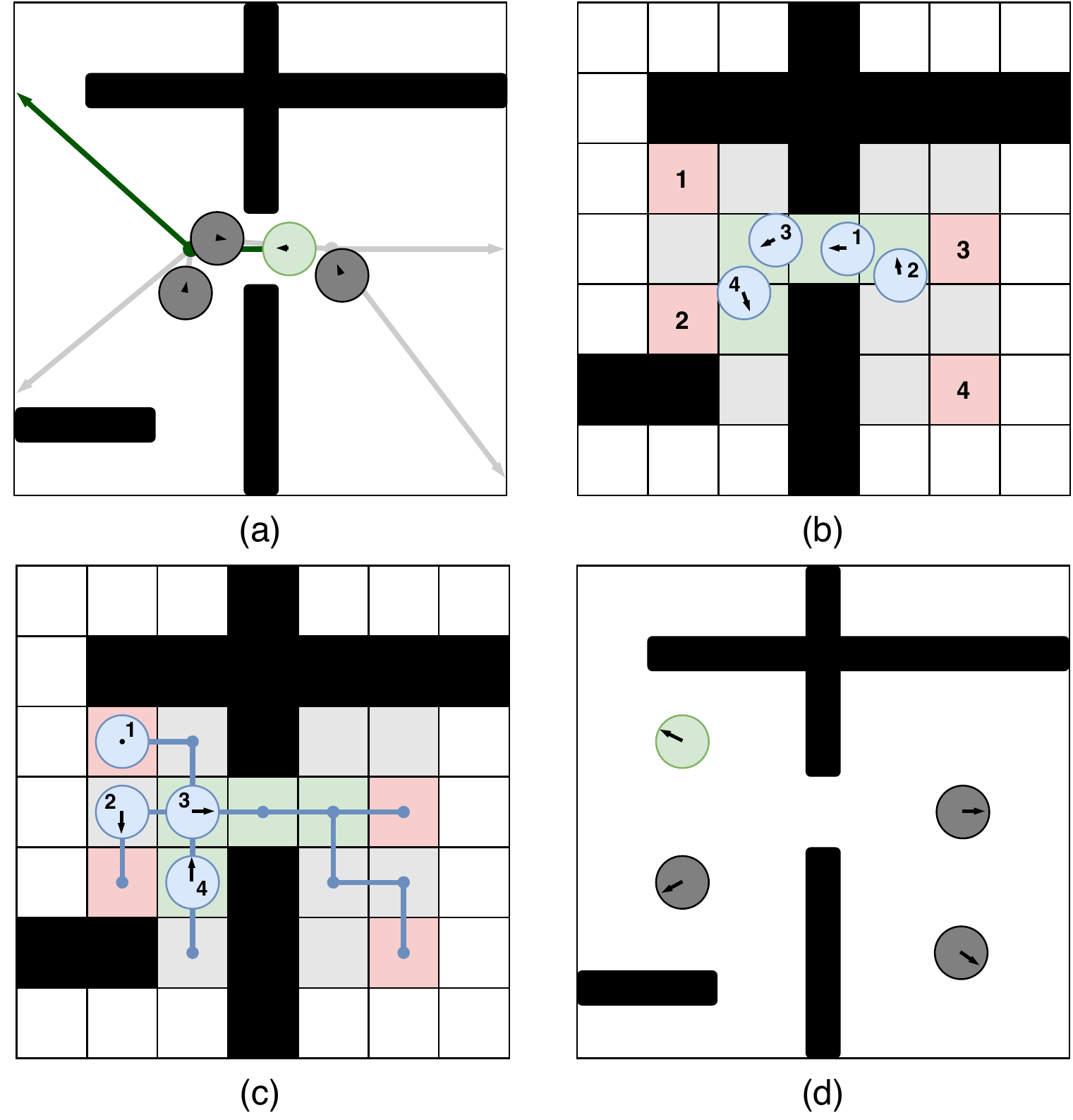} 
    \caption{Creating and solving a MAPF instance upon the deadlock is discovered. Video is available at: \texttt{youtu.be/jE2YrXQ\_pNs}}
    \label{fig:mapf_area}
\end{figure}

When no central controller is present and agents' observation and communication capabilities are limited, each of them typically plans its path and relies on a collision-avoidance mechanism during execution, e.g. Optimal Reciprocal Collision Avoidance (ORCA)~\cite{van2011reciprocal_n}. In many cases, such techniques work well, i.e. make the agents avoid collisions and progress towards their goals. However, in certain scenarios, involving navigation through confined spaces, such mechanisms are likely to fail. Consider an example depicted on Fig.~\ref{fig:mapf_area}-a. Two groups of agents have to switch sides using a passage. Obviously, the individual (egoistic) plans for each agent have them all try to use the passage, causing a traffic jam around the passage. Agents driven by reactive collision avoidance get stuck in a deadlock. They lower down their velocities to zero to avoid collisions and stop progressing towards the goals. Indeed, to reach the goals the agents need to exhibit a form of coordination as some of them have to yield to the others.

In this paper, we study how including locally centralized multi-agent path finding (MAPF) solvers can mitigate this problem and increase the chance of mission completion, see~Fig.~\ref{fig:mapf_area}-b,c,d. For those solvers to be applied we first design a simple, yet effective deadlock detection mechanism. Upon encountering a deadlock a locally-coordinated distributed multi-agent system is formed, i.e. the agents involved in deadlock exchange the information on their current positions and desired goals thus each agent forms and operates with the same (local) world model. Next, we propose a method to create a grid-based MAPF instance that is consistent with the shared world model and serves as the input to one of the modern graph-based MAPF solvers. Specifically, we suggest using a combination of the two prominent algorithms to solve MAPF: \textsc{Push and Rotate}~\cite{de2013push} and a bounded-suboptimal version of Conflict Based Search (\textsc{ECBS})~\cite{barer2014suboptimal}.  
When the MAPF solution is obtained all agents start following it until all of them reach the MAPF goals. When this happens they switch back to following the individual paths.

This work is based on the preliminary study presented in~\cite{dergachev2020combination}. The main novel contributions are as follows.
\begin{itemize}
    \item We present an original, easy-to-implement, domain-independent deadlock detection mechanism.
    \item We present a novel approach to form a MAPF instance used to resolve the deadlock.
    \item We thoroughly describe these routines and present a complete pseudo-code of the algorithm.
    \item We use a combination of the MAPF solvers that complement each other when solving MAPF.
\end{itemize}

Our experiments show that in cases when tight passages are present in the environment the suggested approach significantly outperforms collision avoidance. The success rate raises from 10-15\% to 95-99\% in certain cases.

\section{Related work}
Two lines of research are most relevant to this work: multi-agent path finding and decentralized collision avoidance.

There exist a variety of MAPF formulations, which differ in assumptions on how conflicts between the agents are defined, how agents behave when reaching their goal locations etc.~\cite{stern2019multi}. Still, the intrinsic assumption, that there exists a centralized controller for all agents, is dominant in MAPF research. This assumption is the key to solve MAPF problems optimally by either employing reduction-based techniques~\cite{surynek2012towards} or search-based ones, e.g. Conflict-Based Search (\textsc{CBS})~\cite{sharon2015conflict} which is widely used nowadays to solve MAPF optimally. There exist numerous enhancements that improve the performance of \textsc{CBS}~\cite{sharon2012meta, boyarski2015icbs}. However, getting optimal solutions for a large number of agents under tight time constraints is still problematic. One way to mitigate this issue is to look for bounded-suboptimal solutions, which can be provided by e.g. \textsc{ECBS}~\cite{barer2014suboptimal}. Another way is to use algorithms that adapt ideas from pebble-motion solvers~\cite{surynek2009novel, de2013push}. These algorithms are extremely fast and scale well to large numbers of agents. However, often the cost of their solutions is high. Prioritized planners are known to enjoy the benefits of both approaches -- they often find close-to-optimal solutions in practice and are notably fast and scalable. However, except for some special cases, they are incomplete~\cite{cap2015prioritized}. 

Reactive collision avoidance is also an actively studied field of research. A wide range of algorithms for collision avoidance exist that are different, mainly, in the underlying assumptions regarding movement constraints, observation/communication limitations, etc. The most common approaches at the moment are the ones that are based on the notion of the velocity obstacles. Velocity obstacles were originally described in~\cite{fiorini1998motion}, and in~\cite{van2008reciprocal} a method of reciprocal collision avoidance based on the construction of reciprocal velocity obstacles (RVO) was presented. After that, the velocity-based methods were continuously modified, the formulations of the problems were supplemented with various movement constraints~\cite{van2011reciprocal, snape2011hybrid}. Among the various modifications, the most developed are the ORCA-based algorithms, which, in contrast to the RVO, can guarantee the absence of collisions under certain mild assumptions~\cite{van2011reciprocal_n}. There are multiple algorithms, based on ORCA, that take into account the accuracy of localisation and kinematic constraints~\cite{snape2014smooth, snape2010smooth, alonso2013optimal}. Another approach to collision avoidance is based on the construction of the Voronoi cells~\cite{zhou2017fast, zhu2019b}. These methods involve quadratic programming, thus they are typically slower than ORCA (which relies on linear programming). On the other hand, they can take into account the high-order dynamics and do not require the knowledge of the neighbouring agent's velocities. Finally, machine learning can be used to generate collision avoidance policies~\cite{chen2017decentralized, long2018towards}.

\section{Problem statement}
Consider $n$ agents operating in a 2-dimensional workspace, $W$, which consists of the free space and the obstacles, $W_{free}$ and $W_{obs}=W \setminus W_{free}$. Time, $T$, is discretized, and in each time step each agent can either wait or move into arbitrary direction by applying the velocity $\mathbf{v}$. The state of the agent $i$ at each time step is defined as $\mathbf{p}^i_{t+1}=\mathbf{p}^i_t + \mathbf{v}^i_t\Delta t$, where $\mathbf{p}^i=(x, y)$ is the position of the reference point of the agent $i$ in the world coordinate frame. The velocity is bounded for each agent: $\| \mathbf{v}^i_t \| \leq V^i_{max}$.

A spatio-temporal path (trajectory) for an agent is, formally, a mapping: $\pi: T \rightarrow W_{free}$. It can also be represented as a sequence of agent's locations at each time step: $\pi=\{\pi_0, \pi_1, ... \}$. In this work we are interested in converging trajectories, i.e. such paths by which an agent reaches a particular location and never moves away from it. %This can be formulated as follows: 
%\[\exists k \in T: \pi_{k+t} = \pi_k \: \forall t>1\] 
Time step, $t_{fin} \in T$, by which an agent reaches its final destination defines the cost of the path: $c(\pi)=t_{fin}$. A path is called feasible if an agent following it does not collide with the obstacles and the velocity constraints are met. Two feasible paths are said to be conflict-free if the agents following them never collide.

The problem now is to find a set of paths (one for each agent) starting at the predefined start locations, ending at the predefined goal locations, s.t. each pair of paths is conflict-free. Or, in other words, the problem is to define control inputs, i.e. velocities, s.t. the resulting paths are pairwise collision-free. In this work, we are not imposing a strict requirement on optimizing the cost of the solution, which can be defined as either \emph{flowtime}, $\sum{c(\pi^i)}$, or makespan, $\max{c(\pi^i)}$. However, the lower-cost solutions are obviously preferable.

In the rest of the paper, we rely on the following assumptions, which are not uncommon in the field.
\begin{enumerate}
    \item Workspace, $W$, is tessellated into a regular grid, $G$, composed of the blocked and un-blocked cells. %The size of each cell is $l \times l$. 
    The grid is static and known to the agents.
    
    \item The agents are homogeneous. Each agent is represented as a disk with a radius of less than half of the cell's size. %The maximum velocity is the same for all agents: $V_{max}$.
    
    \item Start and goal locations for each agent are the centers of the grid cells.
    
    \item Each control is executed perfectly, no stochasticity is present. Thus at each time step, an agent's position is deterministic and the agent always knows this position.
    
    \item Each agent is associated with an observation/communication range, $R$. An agent is able to perfectly recover the states of the other agents, who are located within the range, and exchange information with these agents.
\end{enumerate}

Overall, in this work, we abstract away from localization and communication issues, as well as assume a simplistic kinematic model for the agents, to concentrate on the problem of finding conflict-free trajectories. 

\section{Method}

Each agent starts with finding an individual geometric path to its goal that respects the static obstacles. As the workspace is discretized to a grid any algorithm that can find paths on grids can be used, e.g.  \textsc{A*}~\cite{hart1968formal},  \textsc{JPS}~\cite{harabor2011online},  \textsc{R*}~\cite{likhachev2008r}, etc. In this work, we suggest using any-angle planners as they build shorter paths that contain fewer turns compared to \textsc{A*}. In our experiments, we used  \textsc{Theta*}~\cite{nash2007theta}.

    \begin{algorithm}[t]
        \SetKwInOut{Input}{Input}
        \BlankLine
            \Input{W, G: workspace and its grid representation, \\
            start, goal: start/goal location of the agent,\\
            R: agent's observation/communication range}
        \BlankLine
        
        globalPath $\gets$ FindIndividualPath(G, start, goal)\;
        mode $\gets$ normal\;
        \While{goal is not reached}
        {
            neighbors $\gets$ GetNeighborsData(R)\;
            \If{mode = normal}
            {
                \eIf{AgentInDeadlock(neighbors)}
                {
                    MAPFSolution $\gets$ CreateAndSolveMAPF(W, G, neighbors)\;
                    localPlan $\gets$ ExtractIndividualPlan(MAPFSolution)\;
                    mode $\gets$ moveToMAPFStart\;
                }
                {
                    next $\gets$ ComputeCurGoal(globalPath, W)\;
                    $V_{new}$ $\gets$ ComputeSafeVelocity(next, neighbors, W)\;
                }
            }
            \If{mode = moveToMAPFStart}
            {
                \uIf{IsReached(Start(localPlan)) \textbf{and} AgentsReadyToMAPF(neighbors, MAPFSolution)}
                {
                    mode $\gets$ MAPF\;
                }
                \Else
                {
                    $V_{new}$ $\gets$ ComputeSafeVelocity(Start(localPlan), neighbors, W)\;
                }
            }
            \If{mode = MAPF}
            {
                \uIf{IsReached(Goal(LocalPlan)) \textbf{and} AllAgentsEndMAPF(neighbors, MAPFSolution)}
                {
                    mode $\gets$ normal\;
                    next $\gets$ ComputeCurGoal(globalPath, W)\;
                    $V_{new}$ $\gets$ ComputeSafeVelocity(next, neighbors, W)\;
                }  
                \uElseIf{ExternalAgentDetected(neighbors)}
                {
                    MAPFSolution $\gets$ UpdateMAPF(map, neighbors)\;
                    localPlan $\gets$ CurrentAgentPlan(MAPFSolution)\;
                    mode $\gets$ moveToMAPFStart\;
                    $V_{new}$ $\gets$ ComputeSafeVelocity(Start(localPlan), neighbors, W)\;
                }
                \Else
                {
                    $V_{new}$ $\gets$ VelocityByPlan(localPlan)\;
                }
            }
            ApplyNewControl($V_{new}$)\;
        }
        \caption{Single agent navigation}
        \label{alg:main}
    \end{algorithm}

After the path is built an agent starts following it, i.e. it moves from one waypoint to the other by setting the velocity appropriately. In case it encounters other agents within the observation/communication range collision avoidance procedure is triggered and the velocity is set up by this procedure. In this work, we use \textsc{ORCA}~\cite{van2011reciprocal_n} for collision avoidance. This method is fast to compute and provides guarantees on collision not to happen in a certain time window.

At each time step, an agent gathers  information about the states of the agents that are within the communication/visibility range (neighbours). This data is used not only to choose the velocity but also to detect deadlocks as described below in Section~\ref{sec:deadlock_detection}. If a deadlock is detected an agent initiates switching to the \emph{MAPF mode}. As a result, certain agents enter this mode (see Fig.~\ref{fig:mapf_area}-a). These agents share the information about their states and current goals (waypoints on the geometric paths that they want to reach) so each of them possesses the same local world model. The latter is used to create a MAPF instance and solve it (Fig.~\ref{fig:mapf_area}-b,c). We emphasize that each agent operates individually in the \emph{MAPF mode} and no central controller is introduced. However as the operations in this mode are deterministic and each agent knows the states and goals of other agents, the result of forming a MAPF instance and solving it is the same across all involved agents. Consequently, each agent obtains the same MAPF solution -- a set of collision-free plans. It then extracts its individual plan from this solution and follows it to resolve the deadlock. After al agents finish execution of their MAPF plans they switch back to normal mode, i.e. continue moving to the next waypoint on their geometric path utilizing collision avoidance (Fig.~\ref{fig:mapf_area}-d).

The pseudocode of the suggested method is presented in Alg.~\ref{alg:main}. The names of the functions are self-explanatory, e.g. \texttt{ComputeSaveVelocity} stands for determining the velocity command to be applied (e.g. by \textsc{ORCA}), \texttt{AllAgentsEndMAPF} stands for determining whether all of the agents executing the MAPF plans have reached their MAPF goals, etc. We will explain how we design the two most crucial functions, i.e. the one that implements deadlock detection, \texttt{AgentInDeadlock}, and the one that implements creating and solving a MAPF instance, \texttt{CreateAndSolveMAPF}, later in Sections~\ref{sec:deadlock_detection},~\ref{sec:mapf_formation}. 

Please note that the pseudocode introduces an additional mode -- \emph{moveToMAPFStart mode} (Lines 9, 13, 26) and a procedure to update MAPF instance (Lines 23-27). The former is needed as agents, when switching to \emph{MAPF mode}, are likely to be located not at the centers of the grid cells as they freely move in the workspace when following geometric paths and avoiding collisions. On the other hand, MAPF solvers typically rely on a graph representation of the environment and assume that the MAPF start locations (as well as the goal ones) are tied to the graph vertices (centers of the grid cells in our case). Thus before actually executing the MAPF plans, agents need to reach their MAPF starts (determined by \texttt{CreateAndSolveMAPF} routine). To this end, \emph{moveToMAPFStart mode} is introduced. In this mode, agents move to their MAPF starts using collision avoidance as in the normal mode.

Updating MAPF is needed in case an agent which is not part of the group that executes MAPF approaches the agents from that group. In this case, the interfering agent is added to the group and the process of creating and solving MAPF is restarted.

\subsection{Deadlock Detection}
\label{sec:deadlock_detection}

We suggest the following ad-hoc deadlock detection procedure which is based on simple computations and limited information exchange, yet is very efficient in practice (as our experiments show). At each time step, an agent computes its average speed across the last $k$ time steps, where $k$ is the user-specified parameter. If this speed is below the certain threshold $v_{low}$, which means that the agent does not progress towards its goal, it requests the average speed of the agents that are within its communication range. If any of the neighbours report its mean speed to be lower than $v_{low}$, then the deadlock is considered to be detected. The intuition here is that when at least two neighbouring agents do not progress towards their goals it is likely that the main reason for that is their behaviour.

When implementing the described deadlock detection mechanism special care should be made to the cases when agents start moving after exiting the \emph{MAPF mode}. It might be the case that an agent was occupying the MAPF goal waiting for the other agents to finish their plans (Line 19) with $v=0$. So when it switches to \emph{normal mode} and starts moving his average speed might be lower than $v_{low}$, but this is not a deadlock, obviously. To rule out such cases we suggest artificially setting the reported velocity of an agent in \emph{MAPF} mode to $V_{max}$.

\subsection{Forming MAPF Instance}
\label{sec:mapf_formation}
MAPF solvers typically rely on the graph model of the environment and assume that the start and goal locations of the agents are tied to the graph vertices. Thus, to incorporate a MAPF solver into the navigation pipeline we need to \emph{i}) determine the graph in which the MAPF solution will be sought, \emph{ii}) determine the start vertices, \emph{iii}) determine the goal vertices. Naturally, we also need to determine which agents should switch to \emph{MAPF mode}. Next, we explain how we implement these procedures in more detail.

\paragraph{Identifying MAPF participants} Consider agent $i$ that detects a deadlock. The agents that are switched to \emph{MAPF} mode (via the information exchange) are $N(i)$ and $N(N(i))$, where $N$ stands for the neighbours, i.e. the agents that appear to be within the visibility/communication range. Each agent in the \emph{MAPF mode} is randomly assigned a unique priority.

\begin{figure}
    \centering
    \includegraphics[width=0.75\columnwidth]{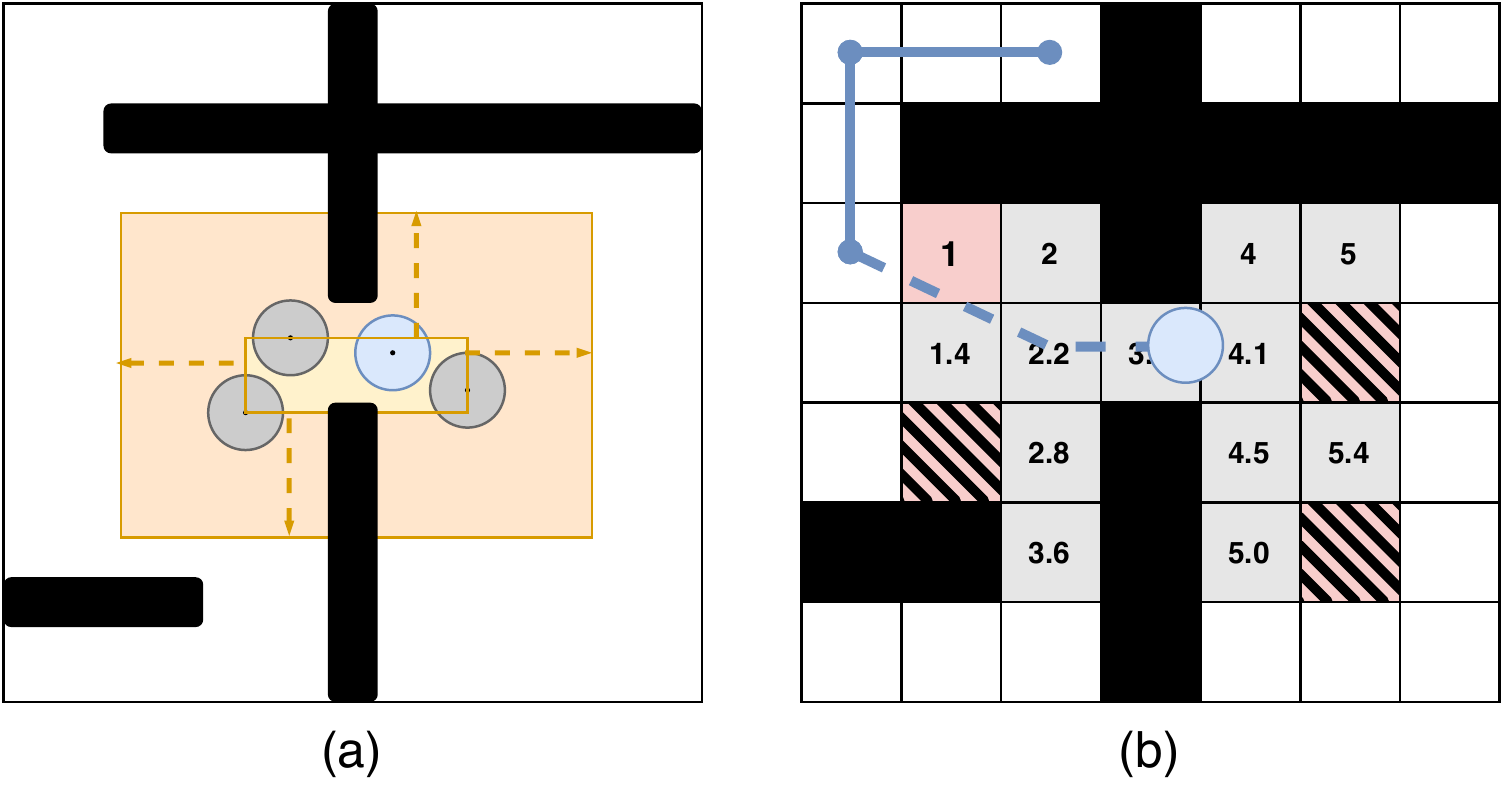} 
    \caption{Constructing a MAPF instance. a) Determining the MAPF area. b) Estimating the MAPF goal for an agent.}
    \label{fig:mapf_construct}
\end{figure}

\paragraph{Determine the graph for MAPF} Recall that the workspace is tessellated to a grid $G$. Naturally, a sub-grid of it, $G'$, is the graph for MAPF. To determine $G'$ the agents share their states $(x, y)$ and identify the minimum and maximum $x$- and $y$-coordinates. These four coordinates form the rectangular area -- see Fig.~\ref{fig:mapf_construct}a. We extend this area by adding the user-defined offset. Finally, we overlap this area with the grid and thus identify which cells from the sub-grid $G'$ to be used in the \emph{MAPF mode}. We assume that agents are allowed to move only between the cardinal cells of $G'$ with the same speed so all moves have a uniform duration. This assumption is common in the MAPF community as it significantly simplifies algorithms' implementation and makes them faster.

\paragraph{Setting start and goal locations} Agents choose start and goal cells in accordance with their priorities. For an agent with priority $l$, we identify the closest un-blocked cell in $G'$ which is not assigned to be the start of the high-priority agents, i.e. $1, 2, ..., l-1$, and set this cell to be the MAPF start for the agent. We denote this cell as $s_{MAPF}$.

To choose the MAPF goal for an agent we identify all cells in $G'$ that are reachable from $s_{MAPF}$ and for each such cell compute the Euclidean distance to the current waypoint on the agent's geometric path, $\pi_{cur}$ (which might be outside $G'$). We now choose a cell that minimizes this distance and, at the same time, does not coincide with a MAPF goal of one of the high-priority agents -- see Fig.~\ref{fig:mapf_construct}-b.

\begin{figure*}[t]
    \centering
    \begin{subfigure}[center]{0.22\textwidth}
        \includegraphics[width=\textwidth]{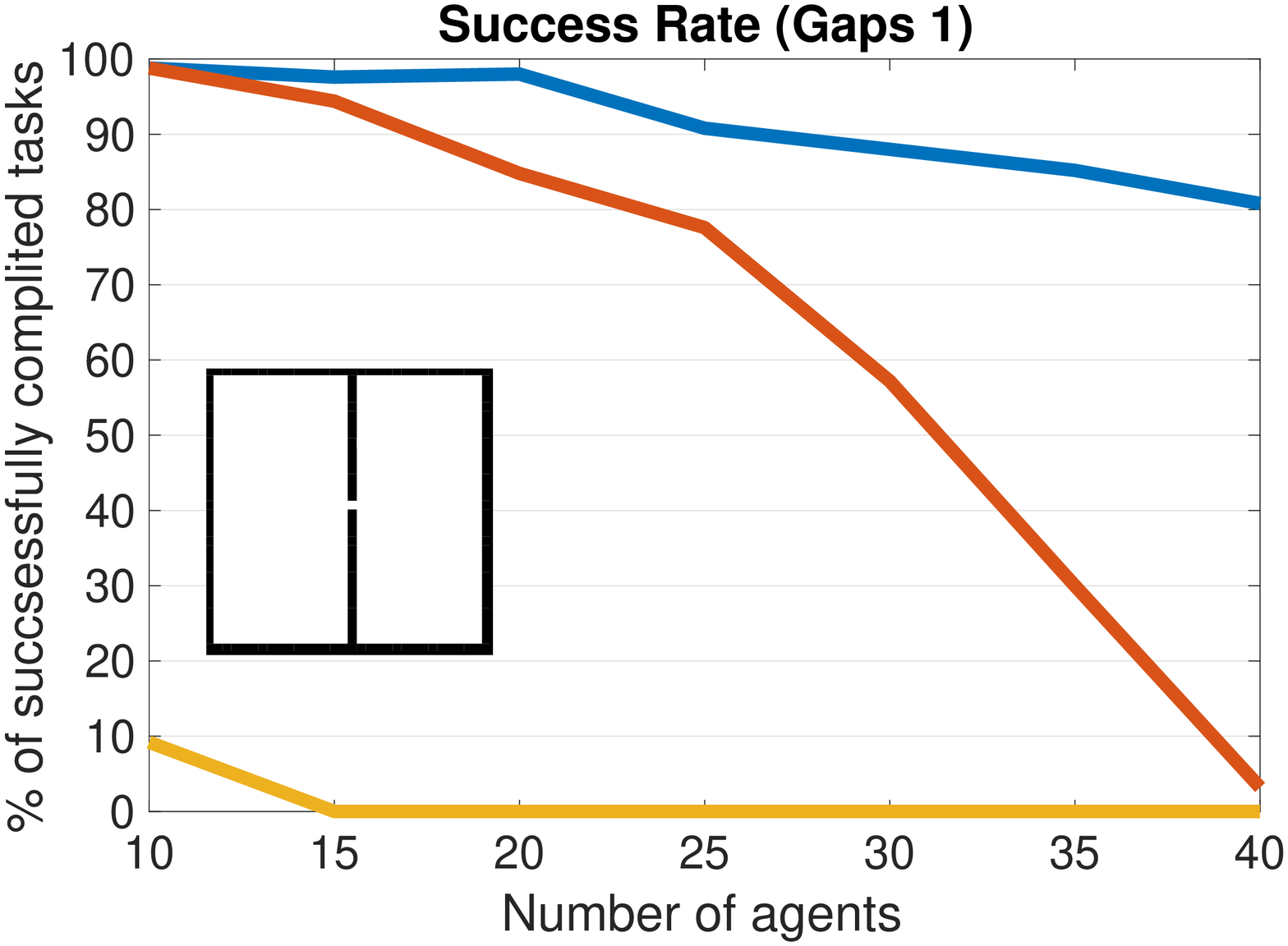} 
        \label{sr_g1}
    \end{subfigure}
    \begin{subfigure}[center]{0.22\textwidth}
        \includegraphics[width=\textwidth]{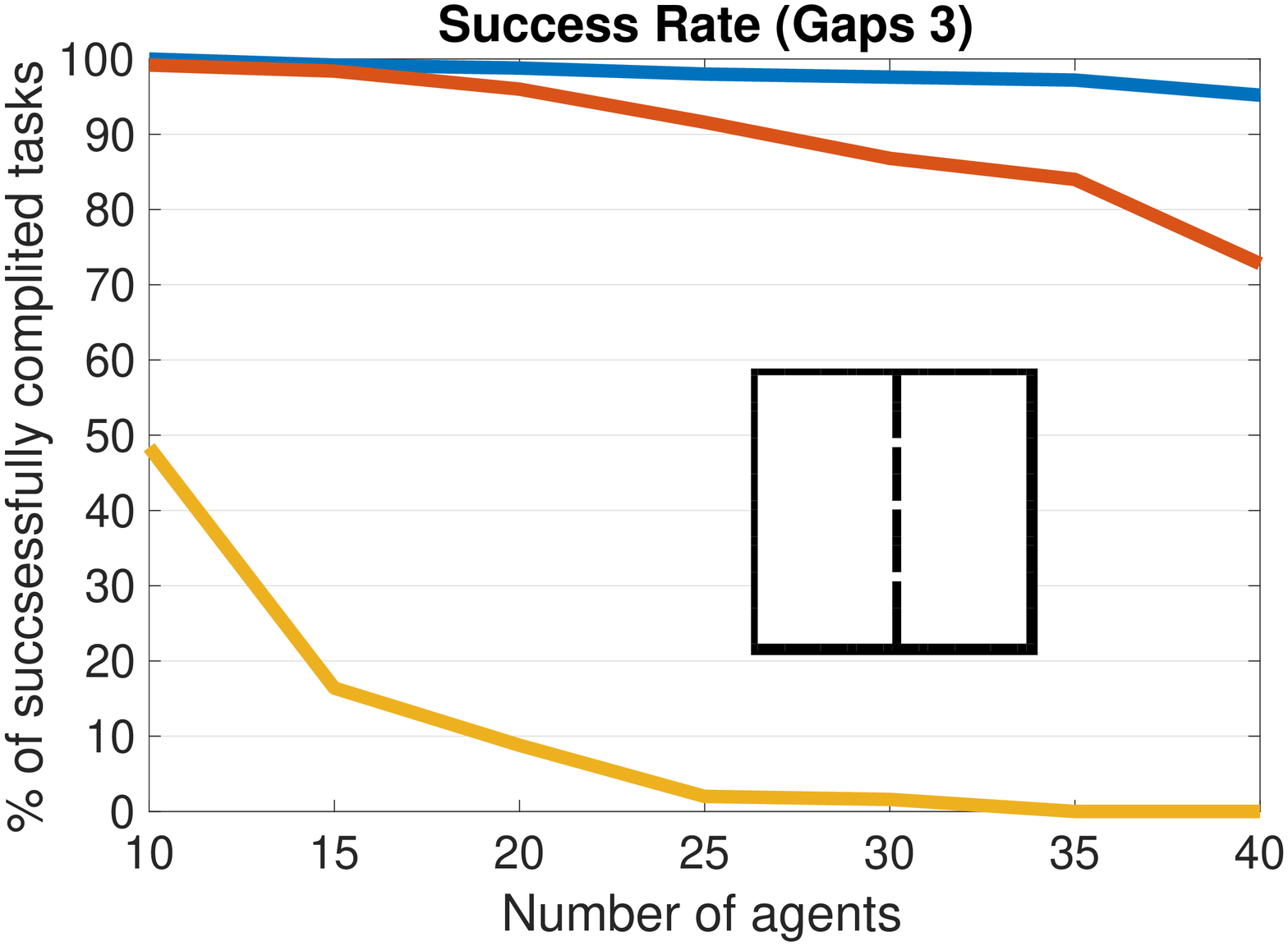} 
        \label{sr_g3}
    \end{subfigure}
    \begin{subfigure}[center]{0.22\textwidth}
        \includegraphics[width=\textwidth]{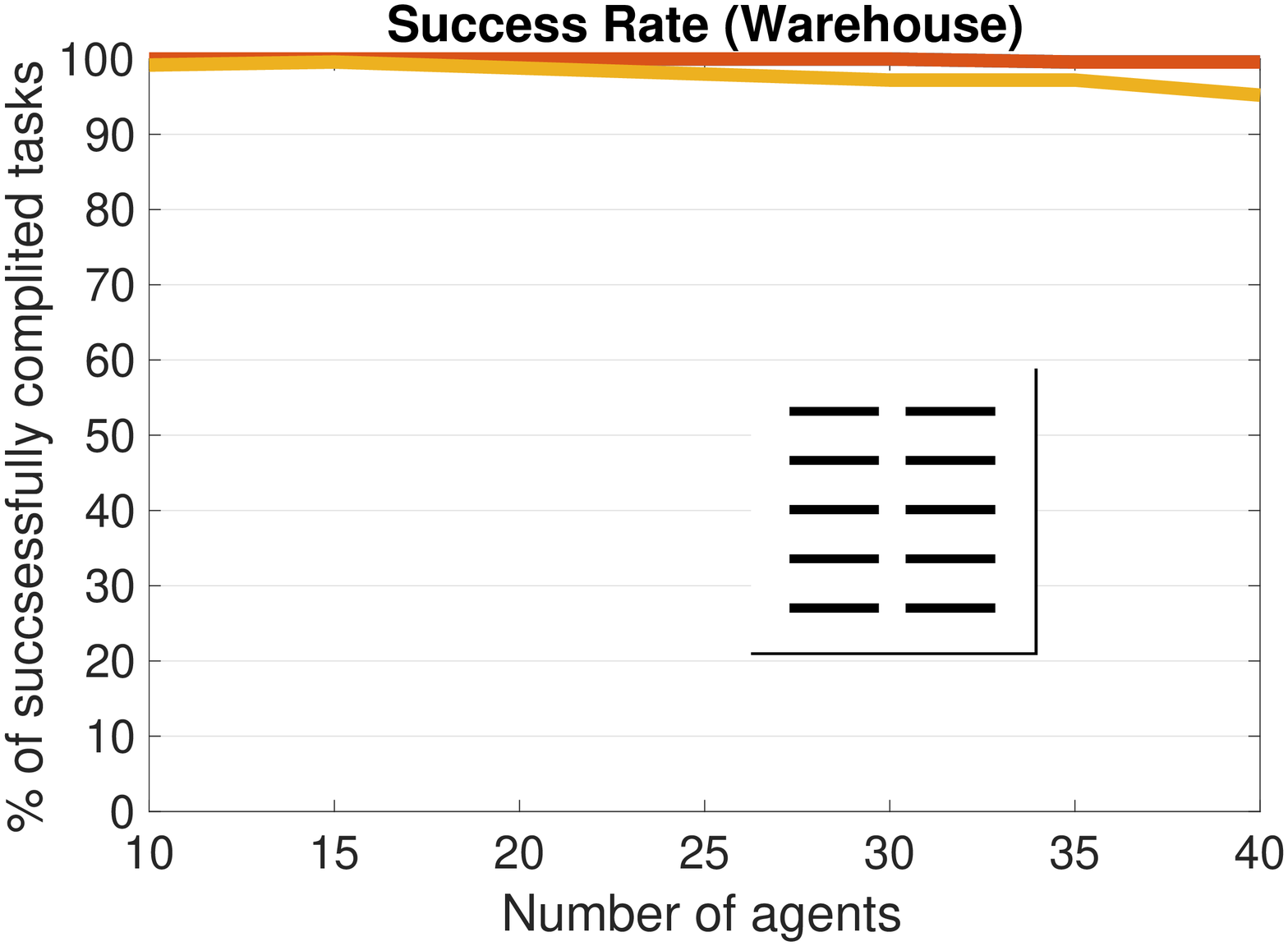} 
        \label{sr_g1}
    \end{subfigure}
    \begin{subfigure}[center]{0.22\textwidth}
        \includegraphics[width=\textwidth]{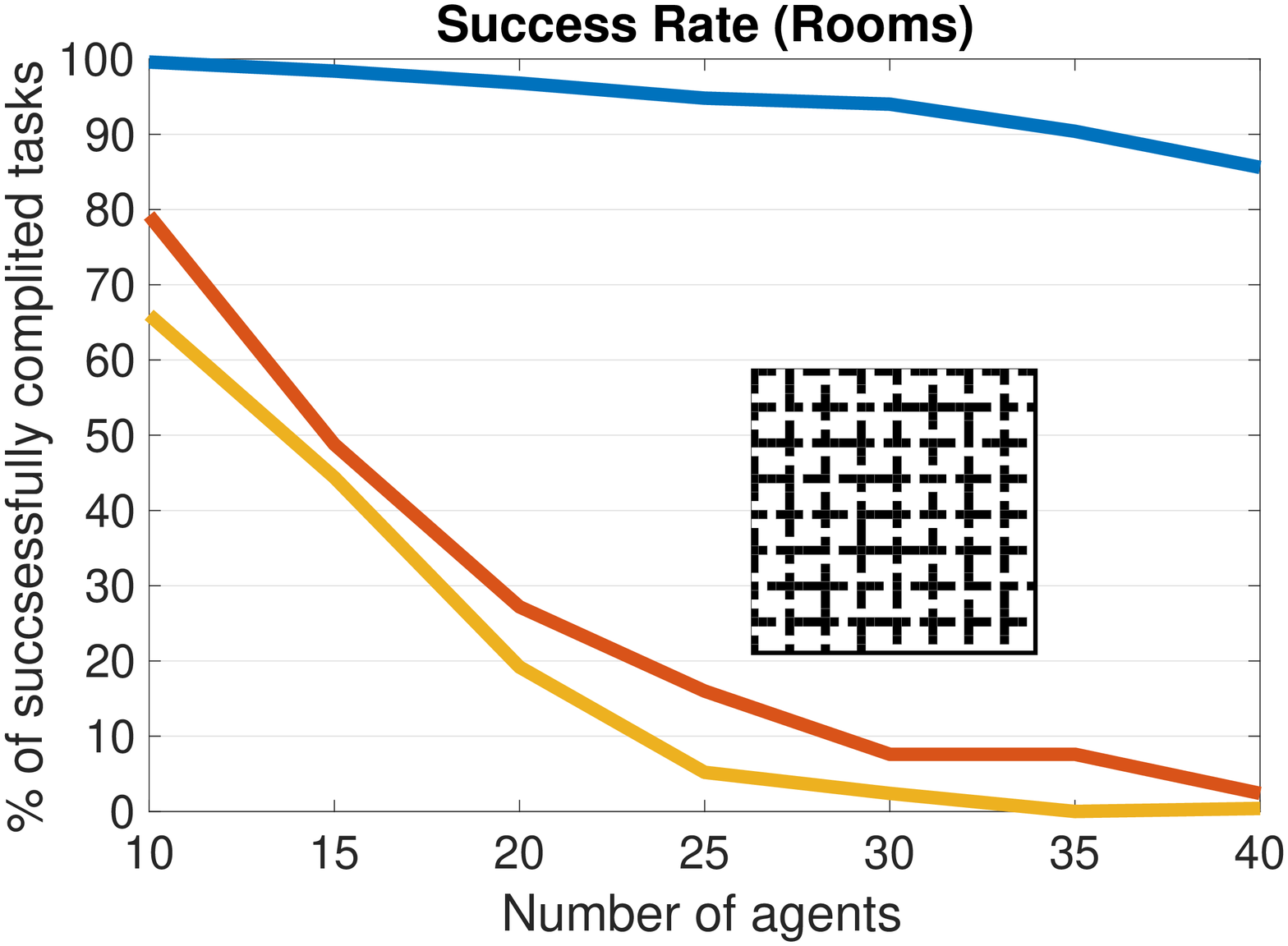} 
        \label{sr_r32}
    \end{subfigure}
    \begin{subfigure}[center]{0.09\textwidth}
        \includegraphics[width=\textwidth]{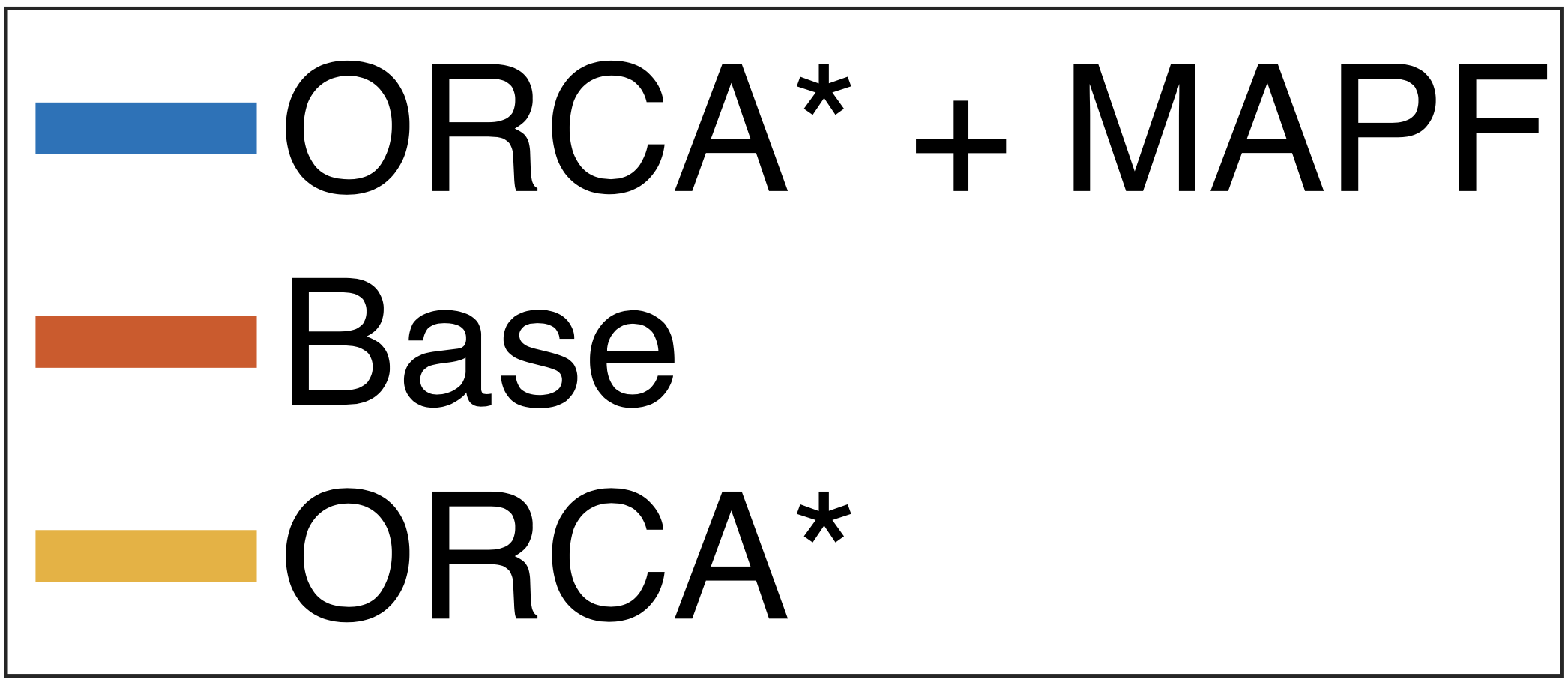} 
        \label{sr_r32}
    \end{subfigure}
    \caption{Success rates of the algorithms on different maps used in the experiments.}
    \label{fig:sr}
\end{figure*}

\begin{figure}
    \centering
    \begin{subfigure}[t]{0.49\columnwidth}
        \includegraphics[width=\textwidth]{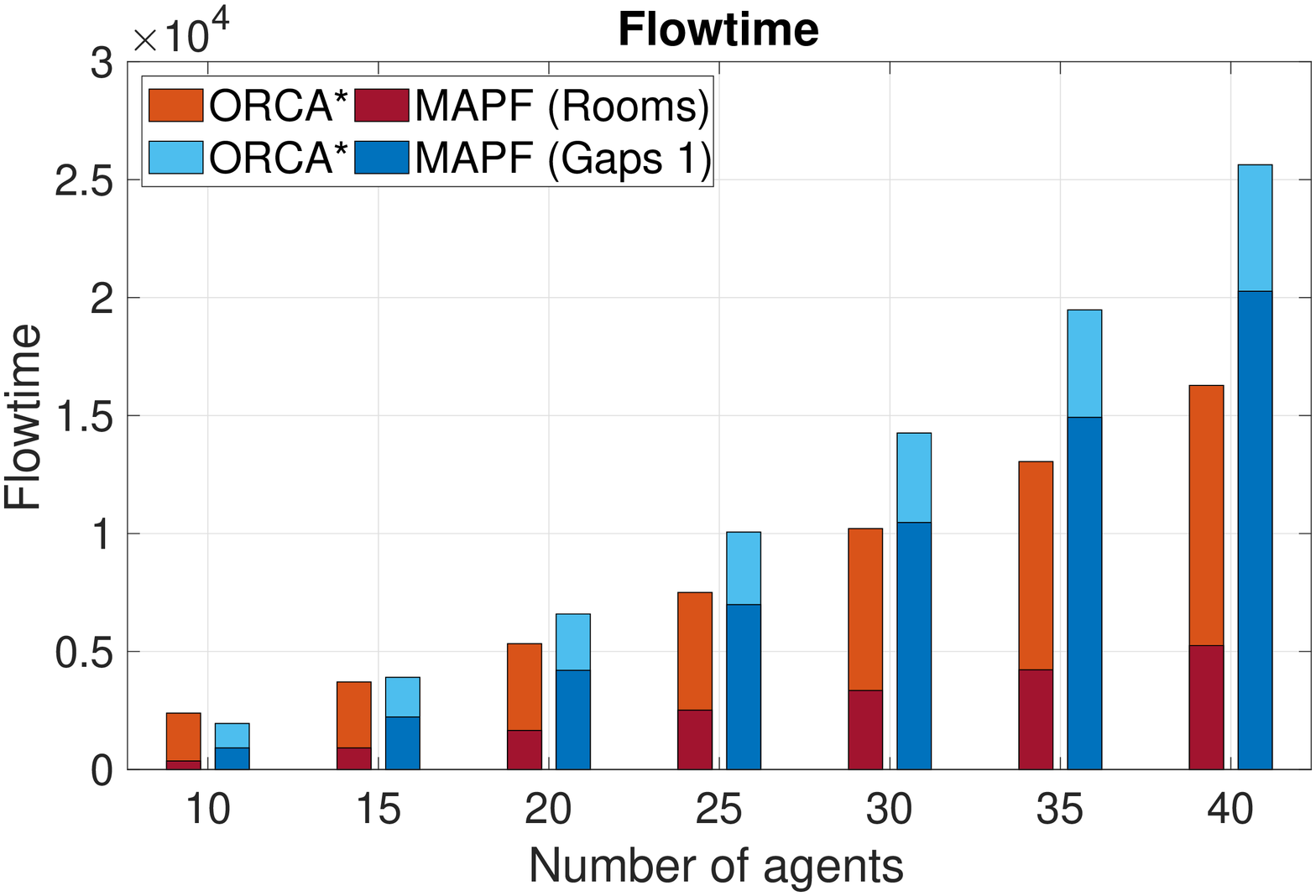} 
        \label{subfig:ft_orca_mapf}
    \end{subfigure}
    \begin{subfigure}[t]{0.49\columnwidth}
        \includegraphics[width=\textwidth]{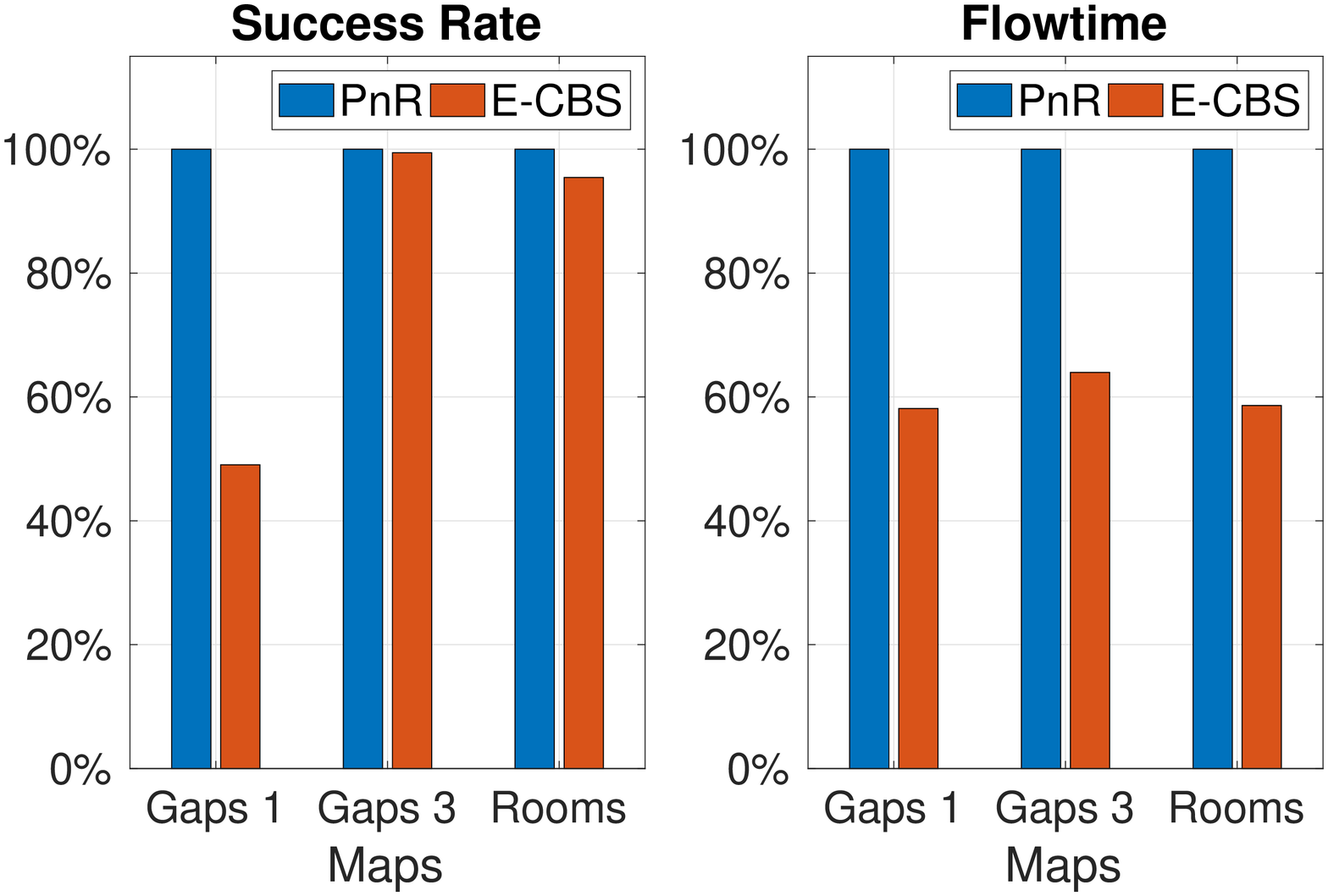} 
        \label{sr_mapf}
    \end{subfigure}
    % \begin{subfigure}[t]{0.32\columnwidth}
    %     \includegraphics[width=\textwidth]{pre_ft_mapf.png} 
    %     \label{ft_mapf}
    % \end{subfigure}
        \caption{Left plot: The break-down between the \emph{MAPF mode} and the normal mode. Two plots on the right: Statistics of the independent MAPF evaluation.}
    \label{fig:results_mapfdetails}
\end{figure}

\subsection{Solving MAPF}

After each agent in the \emph{MAPF mode} has determined the MAPF graph and all the start and goal locations it launches a MAPF solver. Essentially any deterministic MAPF solver can be used to obtain a solution. In this work, we suggest using a combination of \textsc{Push and Rotate}~\cite{de2013push} and \textsc{ECBS}~\cite{barer2014suboptimal} algorithms. The first one is a complete, polynomial-time algorithm by running which we can promptly get a solution. Unfortunately, its cost (both measured as makespan or flowtime) is likely to be high as \textsc{Push and Rotate} by design is not tailored to minimize any cost objective. Thus we suggest invoking a MAPF solver which takes the cost of the solution into account afterwards. We use \textsc{ECBS} for this purpose. This algorithm  guarantees to find bounded-suboptimal solutions. I.e. given a user-specified bound $\epsilon$, it guarantees to return the solution whose cost $c$ satisfies $c \leq \epsilon \cdot c^*$, where $c^*$ is the cost of the optimal solution. Practice-wise we suggest setting up a common time cap for both MAPF solvers and invoke them sequentially. In case, \textsc{Push and Rotate} succeeds in finding a solution but \textsc{ECBS} does not (due to the time limit) we stick to the \textsc{Push and Rotate} solution. In case both algorithms succeed we prefer the solution of \textsc{ECBS}. In the worst case, when \textsc{Push and Rotate} does not succeed we abandon the \emph{MAPF mode} and switch the involved agents back to the normal mode. In this case, the detected deadlock is likely not to be eliminated. In practice, we never encountered the cases when \textsc{Push and Rotate} was not able to provide a solution even when the time limit was set to be 1 second (which is a very strict time limit for a MAPF solver).

\section{Experimental Evaluation}

We implemented the suggested algorithm in C++\footnote{Our code is available at \texttt{github.com/PathPlanning/ ORCA-algorithm}} and evaluated it on a computer equipped with Intel Core i5-8259U (2.3 GHz) CPU with 16 GB of RAM. 

The four different grid maps were used in our experiments. The first two maps are sized $64 \times 64$ and possess a similar structure of two halls separated by a wall with a fixed number of narrow passages. We refer to these maps as \texttt{Gaps~$i$}, where $i$ indicates the number of passages in the wall (either 1 or 3 in our experiments). The third map is sized $64 \times 64$ and contains 10 prolonged rectangular obstacles with plenty of free space between them. We refer to this map as \texttt{Warehouse}. Finally, the fourth map is sized $32 \times 32$ and it represents the indoor environment composed of rooms with passages between them. This map was taken from the \textit{MovingAI}~\cite{stern2019multi} benchmark commonly used in the MAPF community. We refer to this map as to \texttt{Rooms}.

We varied the number of agents from 5 to 40 on each map and for each number of agents created 250 different instances. For the \texttt{Gaps} map, we placed half of the starts in the left hall with the goal in right one and for the other half of the agents -- vice versa. For the other maps, we chose starts and goals randomly. Each agent was considered to be a disk of radius 0.3 of a cell width. When computing the velocities by collision avoidance algorithm this radius was considered to be 0.49 (to further minimize the risk of collision by creating an additional safety buffer around each agent). The observation range of each agent was 3 cells. The maximum speed was 0.1 cells per time step.

We compared the following methods. \textsc{ORCA*} -- a method that combines \textsc{ORCA} collision avoidance with individual path planning by \textsc{Theta*} algorithm. \textsc{ORCA*+MAPF} -- a method that implements the suggested approach. We also included into the comparison a method previously introduced in~\cite{dergachev2020combination} that also combines a MAPF solver and a collision avoidance algorithm but is different in the collision detection and MAPF formation mechanisms as well as it does not use a combination of the MAPF solvers. We refer to this method as \texttt{Base}. The time cap for MAPF part of the \texttt{Base} and \textsc{ORCA*+MAPF} algorithms was 1 second. To set the parameters for \textsc{ORCA*+MAPF} we ran the algorithm on a subset of the generated instances and picked the parameters that resulted in the best performance, i.e. $K=250$, $v_{low}=0.001$ for the deadlock detection and 10 for the sub-optimality factor of \textsc{ECBS}. 

\begin{table}
\centering
\scriptsize

\renewcommand{\arraystretch}{1.0}
\begin{tabular}{c|c c c c c c} 
    % \hline
     & \multicolumn{2}{c}{\texttt{Gaps 1}} & \multicolumn{2}{c}{\texttt{Gaps 3}} & \multicolumn{2}{c}{\texttt{Rooms}} \\
    % \hline
    
     Ag. & $ N_{mapf}$ & $N_{ag}$ & $N_{mapf}$ & $N_{ag} $  & $ N_{mapf} $  & $ N_{ag} $  \\
    \hline
    10 & 3 & 9 & 2 & 4 & 15 & 3\\
    % \hline
    15 & 3 & 13 & 2 & 5 & 30 & 4\\
    % \hline
    20 & 4 & 17 & 5 & 7 & 45 & 5\\
    % \hline
    25 & 5 & 21 & 5 & 9 & 55 & 7\\
    % \hline
    30 & 6 & 24 & 7 & 11 & 66 & 8\\
    % \hline
    35 & 6 & 27 & 8 & 12 & 79 & 8\\
    % \hline
    40 & 6 & 32 & 11 & 17 & 95 & 10\\
    % \hline
    % $\uparrow$  & 100\% & 256\% & 450\% & 325\% & 533\% & 233\%\\
    
\end{tabular}
\caption{MAPF statistics. $N_{mapf}$ stands for the number of MAPF calls, $N_{ag}$ -- for the number of agents involved in MAPF.}
\label{tab:mapf_table}
\end{table}

We ran each algorithm on each instance until either of the conditions met: \emph{i}) all agents reach their goals, \emph{ii}) the average speed of all agents over the last 1000 time steps is below 0.0001, \emph{iii}) time step limit of 20 000 is reached. In the first case, we treat the outcome as a \emph{success}. 

The success rates of the algorithms are presented on Fig.~\ref{fig:sr} (the higher -- the better). The suggested approach clearly outperforms the competitors in all the setups that involve confined spaces and tight passages. E.g. on the \texttt{Gaps 3} map, our approach was able to solve 99\% of tasks involving 40 agents while \textsc{Base}'s success rate was slightly above 70\% and \textsc{ORCA} was not able to solve a single instance in this setup. The reason why our approach has not achieved 100\% success might be due to the transition process that occurs when agents move to their MAPF starts relying only on collision avoidance. The only map for which the difference in success rate is negligible is \texttt{Warehouse}. This is expected as this map contains large portions of the free space and no confined spaces/passages so even basic collision avoidance techniques perform well.

The leftmost plot on Fig.~\ref{fig:results_mapfdetails} shows the averaged flowtime (sum of the trajectories' durations) obtained by our algorithm on the two most challenging maps (\texttt{Rooms} and \texttt{Gaps 1}) and the break-down between the \emph{MAPF mode} and the normal one. As one can see the agents spent more time in the \emph{MAPF} mode on \texttt{Gap 1} map compared to \texttt{Rooms}. A possible explanation is that on this former map the agents quickly approach the passage in normal mode, then find themselves in a deadlock, create and solve MAPF and spend a significant portion of time while following the resultant plans. Upon exiting the \emph{MAPF mode} they appear very close to their goals so the time spent in normal mode while reaching these goals is small again.

We have also analyzed how often the deadlocks were encountered and MAPF solving was triggered. Table~\ref{tab:mapf_table} shows these statistics as well as how many agents were involved in MAPF on average. Evidently, the number of MAPF calls increases with the number of agents. On the \texttt{Rooms} map, this number reaches 95 for 40 agents, while on either of the \texttt{Gaps} maps it always stays below 12. However, the number of agents involved in MAPF is typically higher for the latter two maps and can be up to 32 agents for \texttt{Gaps 1}.

We have additionally analyzed the MAPF instances that occurred in the experiments and separately run \textsc{Push and Rotate} and \textsc{ECBS} on them with the time limit set to 1s (as in the main experiment). The results are shown on Fig.~\ref{fig:results_mapfdetails} on the right. Evidently, \textsc{ECBS} failed to find solutions in numerous cases (which is captured by the success rates plots), however, the cost of its solutions is significantly lower compared to \textsc{Push and Rotate}. Overall, these results confirm that utilizing a combination of planners, as suggested in this work, is an adequate approach that allows combining the strengths of the considered algorithms.

\section{Conclusion}

In this work, we have considered the problem of decentralized multi-agent navigation in confined spaces where deadlocks are likely to occur. We suggested a range of techniques to resolve these deadlocks: deadlock identification, MAPF formation, MAPF solving. We evaluated the presented approach in various scenarios and showed that it clearly outperforms the baseline, i.e. the standard state-of-the-art collision avoidance technique -- \textsc{ORCA}. A prominent direction of future research is to adopt the suggested method to a more challenging problem, e.g. to take into account the kinematic constraints of the agents, to lift the assumption on prior knowledge of the environment, etc.

% \section*{ACKNOWLEDGMENT}

%%%%%%%%%%%%%%%%%%%%%%%%%%%%%%%%%%%%%%%%%%%%%%%%%%%%%%%%%%%%%%%%%%%%%%%%%%%%%%%%

\bibliographystyle{./IEEEtran}
\bibliography{./bibl}

\end{document}